\documentclass[10pt,journal,compsoc]{IEEEtran}
\usepackage{blindtext}
\usepackage{hyperref}

\usepackage{amsmath,amssymb,amsfonts}
\usepackage{hyperref}
\usepackage{enumitem}
\usepackage{algorithmic}
\usepackage{graphicx}
\usepackage{textcomp}

\ifCLASSOPTIONcompsoc
   \usepackage[nocompress]{cite}
\else
    \usepackage{cite}
\fi

\ifCLASSINFOpdf
 
\else
 
\fi

\usepackage{graphicx}
\usepackage{xcolor}

\begin{document}

\title{A Systematic Literature Review on 5G
Security}

\author{\IEEEauthorblockN{ Ishika Sahni\IEEEauthorrefmark{1}  }\\
\IEEEauthorblockA{\textit{\IEEEauthorrefmark{1}School of Engineering,} \\
\textit{University of Guelph, Ontario, Canada  }\\
isahni@uoguelph.ca}
\\
 \and
\IEEEauthorblockN{ Araftoz Kaur\IEEEauthorrefmark{2}}\\
\IEEEauthorblockA{\textit{\IEEEauthorrefmark{2}School of Engineering,} \\
\textit{University of Guelph, Ontario, Canada }\\
araftoz@uoguelph.ca}}




\IEEEtitleabstractindextext{%
\begin{abstract}
\textcolor{black}{{It is expected that the creation of next-generation wireless networks would result in the availability of high-speed and low-latency connectivity for every part of our life. As a result, it is important that the network is secure. The network's security environment has grown more complicated as a result of the growing number of devices and the diversity of services that 5G will provide. This is why it is important that the development of effective security solutions is carried out early. Our findings of this review have revealed the various directions that will be pursued in the development of next-generation wireless networks. Some of these include the use of Artificial Intelligence and Software Defined Mobile Networks. The threat environment for 5G networks, security weaknesses in the new technology paradigms that 5G will embrace, and provided solutions presented in the key studies in the field of 5G cyber security are all described in this systematic literature review for prospective researchers. Future research directions to protect wireless networks beyond 5G are also covered.}}
\end{abstract}

}

\maketitle

\IEEEdisplaynontitleabstractindextext

\IEEEpeerreviewmaketitle

\section{Introduction}\label{sec:introduction}
The 5G wireless system is designed to provide various services and features over the 4G cellular networks. It is intended to deliver various features and functions over the next generation of mobile telecommunications \cite{a1}, \cite{a2}. This evolution of the existing networks is expected to bring about various benefits. One of the main advantages of 5G is its ability to provide enhanced performance and capacity, which will allow it to support massive machine-to-machine communications \cite{a3,a1}. It will also increase the number of people using mobile broadband \cite{a4}.
\\5G aims to reduce the energy consumption and latency of networks by implementing various advanced features \cite{a4}. One of the most important factors that 5G will bring to the table is its ability to provide a 1-millisecond latency. It will also allow it to support a thousand increase in bandwidth per unit area \cite{a5}. Various technologies are being used to implement these features in 5G systems. Some of these include the use of multi-input multiple-output (MIMO), D2D communications,
 and software-defined networks \cite{a6}.
\\The standardization of 5G wireless systems is still in its early stages. There are various use cases that will benefit from this technology, such as industrial automation, smart cities \cite{a3}. It is widely assumed that the vast IoT will enable 5G wireless networks to improve mobile broadband \cite{a7}, \cite{a8}. However, the security and privacy protection of this technology will be very important.
\\Despite the various advantages of 5G, it still has a number of security risks that need to be considered \cite{a6}. Some of these are related to the network itself. However, these also include the connected devices that are used to implement this technology \cite{a9}. For instance:
\begin{itemize}
\item In Canada, the 5G wireless system can be exploited by attackers to prevent the country's 4G and LTE networks from working properly. This vulnerability could allow them to take over the Internet of Things \cite{a6}.
\item Canadian government is actively investigating the security issues of certain suppliers because of the rise in connected devices and the complexity of the 5G infrastructure \cite{a7}.
\item The complexity of the 5G infrastructure and the rise in connected devices will put pressure on networks' security monitoring  \cite{a9}.
\item The evolution of the 5G network from traditional core networks to edge computing is expected to create a huge increase in the number of devices that can connect to the network. However, this also increases the risk of unauthorized access to the network \cite{a10}.
\item Unfortunately, it is not always feasible to provide security features designed to protect the privacy of the data that is collected and stored in cellular networks. There are various issues that can occur in these networks\cite{a10}.
 \end{itemize}

\begin{figure}[h]
	    \centering
	    \includegraphics[width=0.9\linewidth,keepaspectratio]{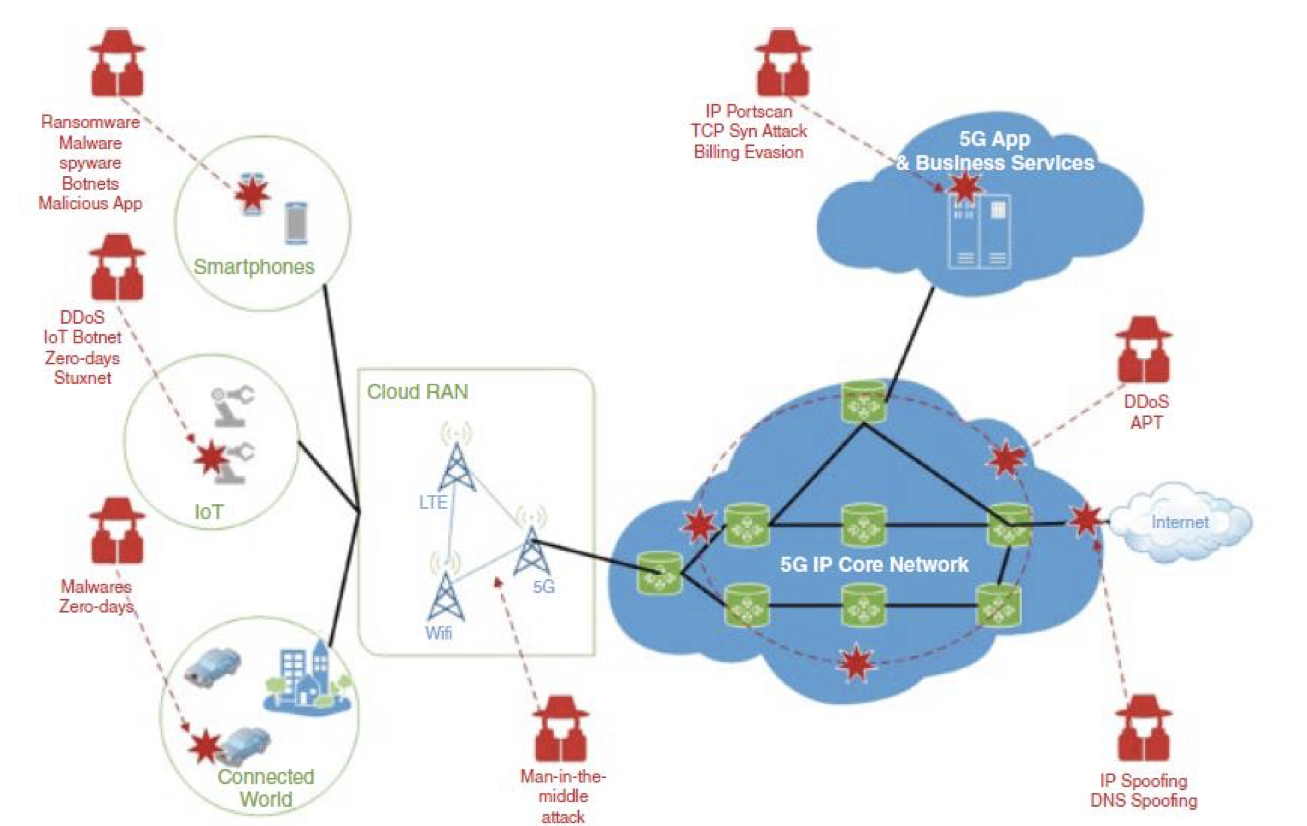}
	    \caption{Privacy Incorporated in the Design of Cloud Computing Environments}
	    \label{MyFig1}
\end{figure}

One of the most common factors that can affect the security of the data that is stored and accessed in cellular networks is the implementation of a traditional security architecture \cite{a11}. Multiple authentications and authorization procedures are included in this type of security feature in order to safeguard the network's integrity \cite{a13}.
\\Figure \ref{MyFig1} depicts the 5G security threat landscape and in addition to using traffic encryption, a base station and a user's device can also be mutually authenticated. This ensures that the security of the mobility management and access of LTE is protected \cite{a14}. Additionally employed to guarantee network securities are a key hierarchy and a handover management technique. It is anticipated that the security requirements for the cellular networks of the future will differ from those of the current generation. The network's security needs must alter in order to conform to the many elements of the 5G network \cite{a15}.

Despite the various security measures that are currently being implemented in 5G networks, the lack of impressive solutions for handling massive connections is still a concern. We believe that there is potential scope for developing security awareness protocols that can address various aspects of data transmission and storage \cite{a5}.
\\In addition, massive MIMO is considered a promising technique for preventing passive eavesdropping \cite{a6}. It can also be used in various applications, such as surveillance, which increases privacy concerns. Because of this, the development of 5G networks can be tied more closely to the services that are tied to it \cite{a7}. For instance, if a major network is down, internet access, TV, and fixed telephone lines can be interrupted simultaneously. To ensure that the 5G system is resilient against security attacks, security automation is needed \cite{a9}. The increasing number of connections of objects and devices in the 5G network has raised concerns about security. Due to the complexity of the network, the number of attacks and vulnerabilities is becoming more evident. This is why it is important that security automation systems are designed to protect the different layers of the network \cite{a11}. 
\\The goal of this project is to contribute to the development of 5G wireless security by analyzing and reporting on the various security issues that affect the availability of information and the integrity of the network. Through our studies, we will also be able to identify new directions in the field.

\subsection{Prior research}
The security of networks is a critical component of the telecommunications industry, which is required to handle the various applications that will be using 5G. These include the Internet of Things (IoT), Digital forensics, and IDS \cite{a11}, \cite{a12}. Due to the increasing number of people using 5G and the complexity of the security challenges that it poses, it is important that the industry takes immediate action to address these issues \cite{a14}. Due to the increasing number of people using 5G and the complexity of the security challenges that it poses, it is important that the industry takes immediate action to address these issues. Before 5G was introduced, the previous generation of networks was mainly concerned with data speed, latency, and crowd movements. However, with the emergence of 6G and the increasing number of people using 5G, there has been a lack of systematic literature reviews on the security of networks \cite{a15}.
\\One of the recent systematic literature reviews by Torres-Carrin \cite{a16}, which was adapted from Bacca and Kitchenham \cite{a17}, was conducted in order to provide a comprehensive analysis of the various aspects of the security of networks. The study focuses on the planning and reviewing of the security of networks, as well as the various vulnerabilities and privacy issues that are affecting the operations of the networks.
\\Q. Tang \cite{a9}, the paper's author, discusses the numerous features of 5G as well as network security. He also talked about how the Internet of Things' problems could be resolved through network technologies (IoT). The study's objective is to develop a methodology that may be used to protect the LTE advanced network.
\\Author Ahmad I. analyzed the various forms of attacks and security challenges that are affecting the operations of networks. He also talked about the various security standards that are being used by the industry to address these issues. Some of these include the 3GPP, 5GPPP, and the Internet Engineering Task Force \cite{a5}.
\\Followed by the study “5G Security Challenges and Opportunities” by A. Dutta. \cite{a17}, the author discussed the various security aspects of the technology and its future directions. He focused on the threat landscape and the evolution of the security standards that are being used by the industry. He additionally talked about the various aspects of the security of the technology and its future directions.
\\Several studies have been conducted on the security aspects of 5G. These studies have shown that the research on technology has increased over the years. It is also expected that the research on the security of networks will continue to grow in 2022 \cite{a19}.

\subsection{Research goals}
The goal of this research is to evaluate the existing research and literature on 5G cyber security and its applications and to summarize the findings of the research in order to provide a more focused view of future developments in this field as shown in Table \ref{Mytable1}.
\begin{table}[h!]
\caption{Research questions.} 
\label{Mytable1}
\setlength{\tabcolsep}{3pt}
\begin{tabular}{p{106pt} p{109pt} }
\hline
 Research Questions(RQ) & Discussion\\
 \hline
 \vspace{0.15 mm} \textbf{RQ1:} What are 5G security applications? 
  & \vspace{0.15 mm}The applications of 5G security will be reviewed to check 5G technologies impact with respect to cyber security.
\\
 \vspace{0.15 mm}  \textbf{RQ2:} What are approaches to improve 5G security?
 & \vspace{0.15 mm}The different methodologies can be deployed related to enhance security of 5G network that can provide better digital infrastructure for security of users and devices. \\
 \vspace{0.15 mm}  \textbf{RQ3:} What are the methods available to improve 5G security?
 & \vspace{0.15 mm} This research question will be analysed to check what methods work best with respect to its current availability.\\
 \hline
\end{tabular}
\end{table}
\subsection{Contribution and Layouts}
The following contributions are provided by this systematic literature study for those who are continuing to watch 5G security, networks, and cyber security and seeking to advance their task:
\begin{itemize}
\item We identify 41 primary studies during the search of 5G security and cyber security papers in the specific field.
\item We analyze the data collected by the various studies and present the findings to provide an updated view of the 5G cyber security field. Through the review, we hope to gain a deeper understanding of this subject.
\item The goal of this review is to examine the several strategies that can be used to increase the safety of various cyber technologies.
\item To encourage additional research in this field, we establish guidelines and make representations.
\end{itemize}

Table \ref{Mytable1} summarizes the three relevant research questions and provides a discussion section where relevant topics can be discussed. The discussion section will also cover the various aspects of 5G security.
\\The format of this essay is as follows: The techniques used to choose the primary studies for analysis in a methodical manner are described in Section 2. The results of all the primary research chosen are presented in Section 3. The findings in relation to the earlier-presented study questions are discussed in Section 4. The research is concluded in Section 5, which also makes some recommendations for additional study.

\section{Research Methodology}
The goal of this review was to find answers to the three research questions presented in  Table \ref{Mytable1}. The guidelines for performing this review were originally created by Kitchenham and were last updated by Chaters and Kitchenham \cite{a17}. The researchers divided the review into three phases. The first one involved planning the review, the second one involved conducting the review, and the third one involved reporting the findings.

\subsection{Selection of primary studies}
The primary studies were gathered from a variety of sources, including journals, digital libraries, references, "grey literature," and the internet.
\\The data was extracted using the boolean operator AND and OR, which were used as search strings:
\\The boolean operators are as follows for IEEE: \textit{“5G” AND (“ network” OR “security”)}.
\\Based on INSPEC the search strings are:\textit{ “5G” AND “mobile” AND “ communication” AND (“networking”  OR “security”)}.

\vspace{1.5 mm}The digital library platforms used during the research are as follows:
\\- IEEE Xplore Digital Library 
\\- SpringerLink 
\\- Google Scholar
\\- UOG Library
\\- ScienceDirect
\\We also searched for articles about the 5G network in addition to these blogs. The description, keywords, or abstract were used in the searches depending on the search platforms. Since September 21, 2022, searches have been conducted.
\\Through the exclusion and inclusion criteria, the results of the searches were filtered. The findings that were generated through the criterion were then analyzed using the method known as snowballing \cite{a15}. This process can be utilized to analyze the literature.

\subsection{Inclusion and Exclusion criteria}
The goal of this article is to provide a review of the literature on cybersecurity in 5G technology, with a particular focus on cybersecurity issues and solutions in this setting. The following research questions were posed for this study:
\\RQ1: What are 5G security applications? 
\\RQ2:  What are approaches to improve 5G security?
\\RQ3: What are the methods available to improve 5G security?
\\The study selection criteria were used to determine the most relevant papers for the review. The results of the search were then analyzed using Google scholar to find the most relevant papers. Table \ref{Mytable2}  illustrates the criteria for the inclusion and exclusion of the papers for the primary studies.
\subsection{Selection results}
Thousands of results related to 5G security were generated by search engines and digital libraries. After analyzing these results, many of them were found to be irrelevant and duplicates, and the results were then analyzed using the inclusion and exclusion criteria. Out of the many best publications that were analyzed, almost half of them were used to interpret the results of the review.

\subsection{ Quality assessment }
The guidelines of Kitchenham and Charters \cite{a17} were followed when it came to the evaluation of the quality of primary studies. They considered various factors, such as the external and internal validity of the data, research bias, and data manipulation. The process was carried out according to the guidelines of Hosseini et al. About 6 to 7 papers were evaluated for quality. The various stages of the process were then carried out to check the reliability of the paper.
\begin{table}[h!]
\caption{Inclusion and exclusion for primary studies} 
\label{Mytable2}
\setlength{\tabcolsep}{3pt}
\begin{tabular}{p{106pt} p{109pt} }
\hline
 Criteria for Inclusion & Criteria for exclusion\\
\hline
 \vspace{0.15 mm}Papers focusing on 5G network and applications  & \vspace{0.15 mm}Paper focusing on the public, economic, and business impact on society\\
 \vspace{0.15 mm}Paper with information to improve 5G security applicable in real-world & \vspace{0.15 mm}Grey literature such as newspapers, blogs, columns \\
 \vspace{0.15 mm}The peer-reviewed journals provided with proper research & \vspace{0.15 mm} Non-English papers or language to known by author\\
\hline
\end{tabular}
\end{table}

The stages of finding the excluded studies, as seen in Table \ref{Mytable3}, are as follows:
\\Stage 1: \textbf{5G security.} The paper must explain what a 5G security network is and its applications. The content should be well-stated and related to resolving the problems with this specific issue.
\\Stage 2:\textbf{Context.} The data should be sufficient to offer the necessary details or observations linked to the study and to give a precise interpretation of the subject.
\\Stage 3: \textbf{5G network applications.} To adequately answer research questions RQ1 and RQ2, the study must contain enough data to demonstrate how the technology or apps have been used to address a specific issue and improve performance.
\\Stage 4: \textbf{Security context.} In order to help in answering RQ3, the paper must explain the security issue and effective, available, or currently available, ways to enhance 5G security.
\\Stage 5: \textbf{5G network performance.} Assessing the performance of 5G security in relation to the strategies used to enhance it and contrasting the various 5G network applications.
\\Stage 6: \textbf{Data acquisition of 5G security.} For accuracy, it is important to assess the specifics of the data that has been obtained.

\subsection{ Data extraction }

\begin{figure}[h]
	    \centering
	    \includegraphics[width=0.9\linewidth,keepaspectratio]{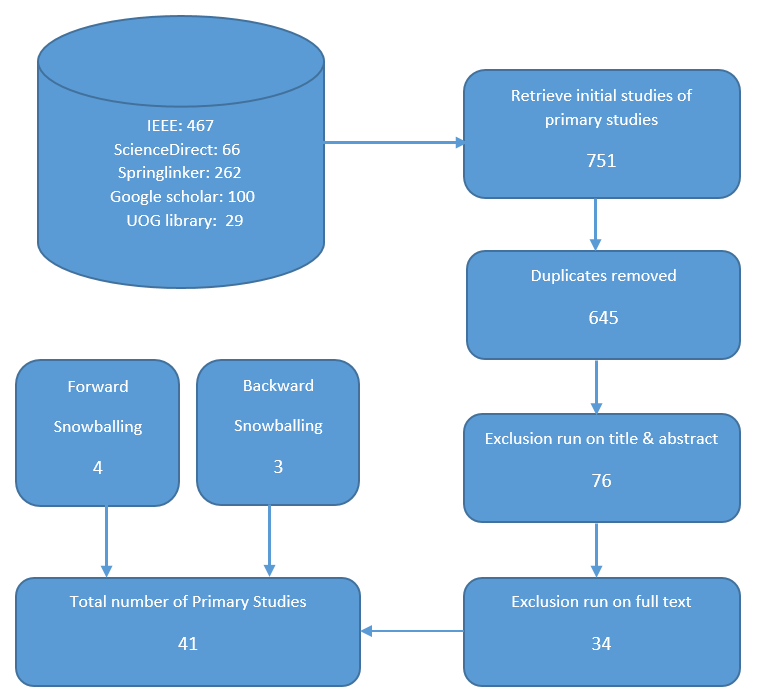}
	    \caption{Privacy Incorporated in the Design of Cloud Computing Environments}
	    \label{MyFig2}
\end{figure}

The objective of this stage is to create data-gathering procedures that will precisely document the data that researchers gather from the primary investigations. The technique is used to lessen the likelihood of prejudice, dishonesty, or invalidity. The Figure \ref{MyFig2} shows the papers selection criteria through the process with the keywords used. The data collected for this review came from various publications. The results of the data synthesis process were then analyzed to determine the best paper for the 5G security issue. The paper was selected based on the various activities involved in the data synthesis process as follows:
\\Descriptive data: In order to comprehend the goal of studies and make comparisons between tasks, the element context is studied.
\\Qualitative data: Results and proposals from various researchers.
\\Quantitative data: When used in a study, data are discovered through testing and research. The retrieved results are used to measure and compare the outcomes.
\subsection{Data analysis}
The data was then analyzed to determine the most relevant questions. Through the quantitive and qualitative synthesis techniques, the final data was then integrated to provide the necessary answers.

\subsubsection{Publications over time}
\begin{figure}[h]
	    \centering
	    \includegraphics[width=0.9\linewidth,keepaspectratio]{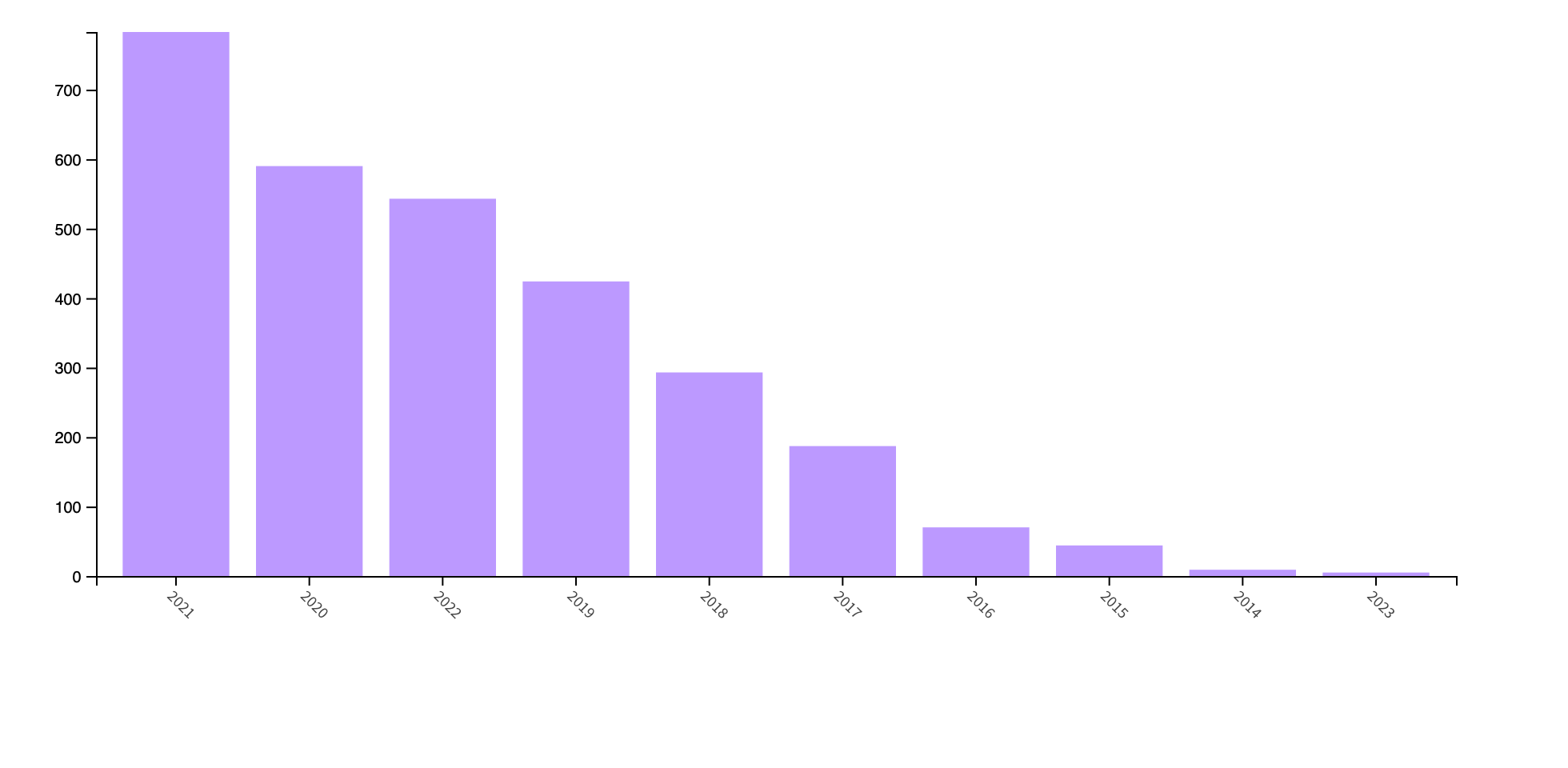}
	    \caption{Privacy Incorporated in the Design of Cloud Computing Environments}
	    \label{MyFig3}
\end{figure}

It has been analyzed that over the years articles and research on the 5G security and network technology have increased and 2021-2022 has the most number of publications as seen in Figure \ref{MyFig3} but it has been analyzed that over the years the publications on the 5G network will reduce as the technology would drift towards more advanced technology that 6G network or more advanced type of network.

\begin{table}[h!]
\caption{Excluded studies} 
\label{Mytable3}
\setlength{\tabcolsep}{3pt}
\begin{tabular}{p{108pt} p{107pt} }
\hline
Checklist for the Criteria Stages& Excluded Studies\\
 \hline
 \vspace{0.15 mm}Stage 1: 5G security  & \vspace{0.15 mm} \cite{s7}\\
 \vspace{0.15 mm}Stage 2: Context & \vspace{0.15 mm} \cite{s3} \\
 \vspace{0.15 mm}Stage 3: 5G network applications & \vspace{0.15 mm} \cite{s11} \cite{s19}\\
 \vspace{0.15 mm}Stage 4: Security context. & \vspace{0.15 mm} \cite{s23}\\
 \vspace{0.15 mm}Stage 5: 5G network performance & \vspace{0.15 mm} \cite{s29} \cite{s33}\\
  \vspace{0.15 mm}Stage 6: Data acquisition of 5G security & \vspace{0.15 mm} \cite{s41}\\
 \hline
\end{tabular}

\end{table}

\subsubsection{Significant keyword counts:}
Relevant keyword counts were used to summarising the recurring themes in the primary studies that were chosen, and a keyword analysis was done across all investigations. Table \ref{Mytab1e4} illustrates the number of keywords during the search of the primary studies. Up to this point, "IOT," "network security," and "5G mobile communication" have been the most often used keywords. In section 3 of the findings, the second-most significant term is further discussed.

\section{Findings}
Each source survey literature was thoroughly studied before pertinent information was extracted and presented in Table \ref{Mytable5} and Table \ref{Mytable6}. There was a lot of talk about the security of 5G, and every major study had a theme or focus relating to how it was handling a specific issue. The attention of each study is also listed in Table \ref{Mytable5} and Table \ref{Mytable6} below.

\begin{table}[h!]
\caption{Number of keywords used in primary studies.} 
\label{Mytable4}
\setlength{\tabcolsep}{5pt}
\begin{tabular}{p{125pt} p{90pt} }
\hline
\vspace{0.30 mm}Keywords   & \vspace{0.30 mm} Count\\
 \hline
 \vspace{-0.40 mm}5G   & \vspace{-0.40 mm}2421\\
 \vspace{-1.50 mm} security& \vspace{-1.50 mm} 2042 \\
 \vspace{-1.50 mm}network & \vspace{-1.50 mm}1626\\
 \vspace{-1.50 mm} mobile& \vspace{-1.50 mm} 1064 \\
 \vspace{-1.50 mm}IoT   & \vspace{-1.50 mm}1018\\
 \vspace{-1.50 mm} communication& \vspace{-1.50 mm}  846  \\
 \vspace{-1.50 mm} data privacy  & \vspace{-1.50 mm}737\\
 \vspace{-1.50 mm} information& \vspace{-1.50 mm} 672  \\
 \vspace{-1.50 mm}smart   & \vspace{-1.50 mm}641\\
 \vspace{-1.5 mm} privacy& \vspace{-1.5 mm}  584  \\
 \vspace{-1.5 mm} control  & \vspace{-1.5 mm}575\\
 \vspace{-1.5 mm} Internet& \vspace{-1.5 mm} 551 \\

 \hline
\end{tabular}
\end{table}

The main themes of the primary studies were grouped together to make it easier for the readers to understand their findings. For instance, those that looked into system architecture, technology, and telecommunications were categorized under the network and communication group.
As per the findings, most of the primary studies that are conducted regarding the security of 5G networks focus on the protection of the Internet of Things. The researchers identified the importance of data privacy and security as the next most critical factors. The third most common theme that the researchers discussed was security networking. They were mainly focused on how 5G networking and security can be integrated into the same technology. Network slicing and data privacy were the fourth most common topics that the researchers talked about. The fifth most common theme that the researchers talked about was machine learning and WiFi. They were also focused on how these technologies can be used to solve the security issues related to 5G.
\section{Discussion}
The initial keyword searches for 5G security revealed that there were a lot of papers related to this subject. The evolution of the technology and its security systems is expected to continue with the coming years as 6G and fibers are introduced.
Most of the studies that were selected for publication were experimental concepts or proposals that were designed to solve today's security problems. Some of the solutions that were presented in these studies were very practical. These include methods that are used to solve various security issues related to 5G, privacy, and networks.

\begin{table}[]
\caption{Main findings and themes of the primary studies.} 
\label{Mytable5}
\centering
\tiny

\begin{tabular}{|c|c|c|}
\hline
\multicolumn{1}{|l|}{Primary Studies} & \begin{tabular}[c]{@{}c@{}}Key Qualitative and Quantitative Data\\ Reported\end{tabular}                                                                                                                                                                                                   & \begin{tabular}[c]{@{}c@{}}Types of Security \\ Applications\end{tabular}          \\ \hline
\cite{s1}                                    & \begin{tabular}[c]{@{}c@{}}The UbiPriSEQ framework uses Deep \\ Reinforcement Learning (DRL) to \\ maximize security, privacy, and threats\\ in 5G IoT in an adaptive, dynamic, and \\ comprehensive manner.\end{tabular}                                                                  & IoT                                                                                \\ \hline
\cite{s2}                                    & \begin{tabular}[c]{@{}c@{}}With the implementation and use of\\ 5G-based security systems, work is\\ being done to enable action to be taken\\ on the security implications \\ and cyber dangers.\end{tabular}                                                                             & Security                                                                           \\ \hline
\cite{s4}                                   & \begin{tabular}[c]{@{}c@{}}Work on software-defined networking, \\ the Internet of Things, and large machine-type \\ communication incurred in encoding and \\ decoding for 5G security\end{tabular}                                                                                       & \begin{tabular}[c]{@{}c@{}}IoT or \\ mMTCs\end{tabular}                            \\ \hline
\cite{s5}                                    & \begin{tabular}[c]{@{}c@{}}To perform data security in 5G networks, \\ a number of encryption algorithms \\ and cryptanalysis techniques have \\ been mentioned.\end{tabular}                                                                                                              & \begin{tabular}[c]{@{}c@{}}Data \\ Privacy\end{tabular}                            \\ \hline
\cite{s6}                                    & \begin{tabular}[c]{@{}c@{}}This paper aims to discuss the various security\\ risks associated with 5G network slicing. \\ It also explores the recommendations for\\ implementing effective intra-slice and\\ life-cycle security.\end{tabular}                                            & \begin{tabular}[c]{@{}c@{}}Network \\ Slicing\end{tabular}                         \\ \hline
\cite{s8}                                    & \begin{tabular}[c]{@{}c@{}}The paper presents an overview\\ of the various security mechanisms \\ that are required to protect the data that \\ are collected and stored in 5G networks. \\ It also explores the potential risks that\\  can be encountered by these systems.\end{tabular} & Security                                                                           \\ \hline
\cite{s9}                                    & \begin{tabular}[c]{@{}c@{}}An overview of the new ideas in the \\ the architecture of the 5G security\\ core network and their implications for \\ security.\end{tabular}                                                                                                                  & \begin{tabular}[c]{@{}c@{}}Security and \\ Networking\end{tabular}                 \\ \hline
\cite{s10}                                   & \begin{tabular}[c]{@{}c@{}}An evaluation of the security and privacy\\ threats posed by 5G applications, as well \\ as research on associated security standards, \\ is provided, along with recommendations\\ for addressing these risks.\end{tabular}                                    & \begin{tabular}[c]{@{}c@{}}Security and \\ Privacy\end{tabular}                    \\ \hline
\cite{s12}                                   & \begin{tabular}[c]{@{}c@{}}Supporting the development of ML models\\ that contribute to the provision of highly\\ dynamic and reliable security procedures\\ for the software-centric networks of \\ 5G security.\end{tabular}                                                             & \begin{tabular}[c]{@{}c@{}}Machine\\  Learning\end{tabular}                        \\ \hline
\cite{s13}                                   & \begin{tabular}[c]{@{}c@{}}The 5G network architecture’s security\\ specifications are described in general terms, \\ along with user privacy.\end{tabular}                                                                                                                                & \begin{tabular}[c]{@{}c@{}}Network \\ Security\end{tabular}                        \\ \hline
\cite{s14}                                   & \begin{tabular}[c]{@{}c@{}}A thorough investigation of the weaknesses \\ in the devices’ design that allows for\\  eavesdropping and charge avoidance\\  when connected to the 5G network\end{tabular}                                                                                     & IoT                                                                                \\ \hline
\cite{s15}                                   & \begin{tabular}[c]{@{}c@{}}A description of the fundamental methods\\  and existing E2E secure communication\\ scenarios, together with the necessary\\ conditions based on security threat analysis\end{tabular}                                                                          & \begin{tabular}[c]{@{}c@{}}Communication \\ Security\end{tabular}                  \\ \hline
\cite{s16}                                   & \begin{tabular}[c]{@{}c@{}}The paper also explores the various \\ security threats that can be encountered\\  in the architecture and operations of\\ 5G-LPWAN-IoT and smart cities.\end{tabular}                                                                                          & IoT                                                                                \\ \hline
\cite{s17}                                   & \begin{tabular}[c]{@{}c@{}}The paper explores the various security\\ issues that are related to the 5G-V2X \\ communication levels. It aims to develop a \\ strategy that will allow the security of the \\ 5G-V2X ecosystem to be handled using \\ blockchain technology.\end{tabular}    & \begin{tabular}[c]{@{}c@{}}Communication\\  Security\end{tabular}                  \\ \hline
\cite{s18}                                   & \begin{tabular}[c]{@{}c@{}}This paper aims to provide a comprehensive\\ analysis of the security vulnerabilities \\ in the network layer of various wireless \\ sensor networks and the Internet of Things.\end{tabular}                                                                   & IoT                                                                                \\ \hline
\cite{s20}                                   & \begin{tabular}[c]{@{}c@{}}A detailed review of the latest developments\\ in the security of the 5G network \\ and discusses how to secure it while \\ meeting the emerging service requirements.\end{tabular}                                                                             & \begin{tabular}[c]{@{}c@{}}Security and \\ Privacy\end{tabular}                    \\ \hline
\cite{s21}                                   & \begin{tabular}[c]{@{}c@{}}It highlights the importance of 5G technology \\ in intelligent automation and industry \\ digitisation and examines how it affects \\ various manufacturing organi- zations.\end{tabular}                                                                      & IoT                                                                                \\ \hline
\cite{s22}                                   & \begin{tabular}[c]{@{}c@{}}An in-depth understanding of IoT \\ security methods and how they can \\ improve in the future\end{tabular}                                                                                                                                                     & Security                                                                           \\ \hline
\cite{s24}                                   & \begin{tabular}[c]{@{}c@{}}The study examined how IoT technologies\\  and applications are improved by 5G \\ networks. It compares the problems and \\ potential solutions when a 5G network \\ is implemented.\end{tabular}                                                               & IoT                                                                                \\ \hline
\cite{s25}                                   & \begin{tabular}[c]{@{}c@{}}SDN networks are utilized to decrease \\ the leakage of sensitive information, \\ lower computing costs, streamline the \\ encryption process and minimize \\ associated dangers.\end{tabular}                                                                  & \begin{tabular}[c]{@{}c@{}}Data\\  encryption\end{tabular}                         \\ \hline
\cite{s26}                                   & \begin{tabular}[c]{@{}c@{}}Highlights the recent enhancements in \\ 5G technology and techniques used with \\ respect to machine learning\end{tabular}                                                                                                                                     & \begin{tabular}[c]{@{}c@{}}Machine \\ Learning\end{tabular}                        \\ \hline
\cite{s27}                                   & \begin{tabular}[c]{@{}c@{}}This document explains what 5G is and \\ what security precautions can be \\ made to enhance it.\end{tabular}                                                                                                                                                   & \begin{tabular}[c]{@{}c@{}}Network \\ Security\end{tabular}                        \\ \hline
\cite{s28}                                   & \begin{tabular}[c]{@{}c@{}}The article illustrates the significance\\ of 5G data security and privacy \\ as well as how unstructured data\\ may compromise customer privacy.\end{tabular}                                                                                                  & \begin{tabular}[c]{@{}c@{}}Data \\ Privacy\end{tabular}                            \\ \hline
\cite{s30}                                   & \begin{tabular}[c]{@{}c@{}}The paper presents an overview of\\ the various security issues that are \\ related to the 5G environment. It also \\ explores the proposed technologies \\ that can be used to secure it\end{tabular}                                                          & \begin{tabular}[c]{@{}c@{}}Data \\ Security\end{tabular}                           \\ \hline

\end{tabular}
\end{table}

\begin{table}[]
\caption{Main findings and themes of the primary studies.} 
\label{Mytable6}
\centering
\tiny

\begin{tabular}{|c|c|c|}
\hline
\multicolumn{1}{|l|}{Primary Studies} & \begin{tabular}[c]{@{}c@{}}Key Qualitative and Quantitative Data\\ Reported\end{tabular}                                                                                                                                                                    & \begin{tabular}[c]{@{}c@{}}Types of Security \\ Applications\end{tabular}          \\ \hline
\cite{s31}                                   & \begin{tabular}[c]{@{}c@{}}The survey aims to provide an overview\\  of the various factors that can affect\\  the security of the 5G environment. \\ It also explores the possible solutions that \\ can be made to address these issues.\end{tabular}     & \begin{tabular}[c]{@{}c@{}}Network \\ Security\end{tabular}                        \\ \hline
\cite{s32}                                   & \begin{tabular}[c]{@{}c@{}}The study discusses machine learning\\ techniques that can aid in overcoming \\ difficulties in SDN-based 5G networks.\end{tabular}                                                                                              & \begin{tabular}[c]{@{}c@{}}Machine \\ Learning\end{tabular}                        \\ \hline
\cite{s34}                                   & \begin{tabular}[c]{@{}c@{}}The study examines the improvements\\  in 5g security needed for Internet of\\  Things users, emphasizing crucial\\  problems with location-tracking and \\ network degradation.\end{tabular}                                    & IoT                                                                                \\ \hline
\cite{s35}                                   & \begin{tabular}[c]{@{}c@{}}The peer-reviewed journal \\ explores the difficulties encountered\\  and potential solutions to raise \\ privacy, security, and trust.\end{tabular}                                                                             & \begin{tabular}[c]{@{}c@{}}Data \\ Privacy\end{tabular}                            \\ \hline
\cite{s36}                                   & \begin{tabular}[c]{@{}c@{}}The paper discusses the difficulties\\ in conducting research and upcoming\\ projects related to physical layer\\ security in IoT technologies with\\ regard to 5G security.\end{tabular}                                        & IoT                                                                                \\ \hline
\cite{s37}                                   & \begin{tabular}[c]{@{}c@{}}This paper discusses the various\\ steps involved in implementing\\ private and public blockchains in the\\ 5G environment. It also explores the \\ advantages of using blocks as a secure\\ authentication method.\end{tabular} & \begin{tabular}[c]{@{}c@{}}Peer-to-peer \\ network for \\ 5G security\end{tabular} \\ \hline
\cite{s38}                                   & \begin{tabular}[c]{@{}c@{}}In order to increase the security of\\  its design, Software Defined \\ Networks (SDN) and Network Function \\ Virtualization (NFV) technologies \\ provide unique functions\end{tabular}                                        & \begin{tabular}[c]{@{}c@{}}Network \\ Security\end{tabular}                        \\ \hline
\cite{s39}                                   & \begin{tabular}[c]{@{}c@{}}A detailed overview of the\\ various schemes that are \\ currently being used in the \\ 5G environment. These include \\ software-defined networks, \\ heterogeneous networks, and the \\ Internet of Things\end{tabular}        & Security                                                                           \\ \hline
\cite{s40}                                   & \begin{tabular}[c]{@{}c@{}}In the study, security issues \\ related to WiFi and LiFi network \\ interfaces with 5G networks are discussed.\end{tabular}                                                                                                     & WiFi                                                                               \\ \hline
\end{tabular}
\end{table}

In one of the primary studies, "5G Security Innovation with Cisco," the author David R. Geller \cite{s2}  provides a variety of innovative ideas that help you address the security challenges that will be faced by the operators and consumers of next-generation networks. He also emphasizes the importance of evaluating the security risks that will be involved in the development and operation of 5G. The rapid emergence and evolution of 5G services will bring an extension of the existing cyber risk landscape. This is why it is important that the security community is aware of the potential risks that will be involved in the development and operation of these networks \cite{s2}, \cite{s24}.
\newline
On the other hand, the author Olimid's [S6] primary study focused on the security aspects of the 5G network slicing. It highlighted the various threats that will be presented by this technology. The study identified various research directions that can help address the security concerns of 5G network slicing. These include the use of artificial intelligence, end-to-end security, and automated defense mechanisms. Besides these, the study also highlighted the need for rigorous security models for network slicing \cite{s6}.
\newline
Below are discussed in-depth the three research questions for better understanding and information for researchers gathered from the several primary studies listed above.

\subsubsection{RQ1: What are 5G security applications?}
The literature review highlights the importance of incorporating the various applications and functions in 5G security. In order to secure the network layer, it is necessary that the security community has the necessary resources and capabilities to implement bio-inspired security measures. Unfortunately, the current cybersecurity infrastructure is not designed to address the various challenges that will be presented by the emergence and evolution of 5G networks. This is why it is important that the security community has the necessary resources and capabilities to implement bio-inspired security measures. Organizations can increase their efficiency by using the Internet of Things, a technology that offers greater computational and analytical capability. It is a rapidly evolving field that is expected to continue to expand due to the emergence of new innovations such as smart terminals and cloud computing \cite{s18}\cite{s21}. 

\textbf{IoT} - The emergence of 5G-enabled networks and the need for resilient security are some of the factors that have prompted the development of bio-inspired cybersecurity techniques. These are designed to help prevent unauthorized access and use of networks. Some of the techniques that are commonly used in this area include anti-virus, intrusion detection, and threat analysis. Multiple studies have been conducted on the use of bio-inspired techniques in networking systems. Numerous studies have been conducted on the use of bio-inspired techniques in networking systems. These studies have contributed to the advancement of cybersecurity \cite{s18}. 

\textbf{Data storage and sharing }- Due to the availability of certain vulnerabilities in the networks' data transmission paths, attackers can perform various attacks to alter the data transmission reliability of the networks. These assaults may result in a decline in the networks' actual quality \cite{s19}.

\textbf{Network security} - In most studies, the techniques used in these studies are referred to as bio-inspired techniques. In this paper, the author Choras et al. \cite{a20} shows the various applications of these techniques and how they can help improve the security of networks. Researchers are led by various reasons why bio-inspired techniques are commonly used in cybersecurity. These include their ability to adapt to different environmental conditions, resilience to failures, and the ability to self-evolve \cite{s18}.

\textbf{Private user data} - Due to the emergence of 5G, organizations are expected to take more proactive steps in addressing the various regulatory and privacy risks that are associated with the use of this technology. For instance, they are required to implement effective measures to improve transparency and manage the privacy of their customer’s data. In addition, the availability of stored data can be used by various organizations for various purposes, such as analyzing and improving their products and services. This type of data can also be used to develop new and innovative products and services. To ensure that their customers are informed about the use of their data, service providers are required to provide clear and comprehensive information regarding the data's purpose and how it will be used \cite{s20}.

\textbf{Navigation and utility of the World Wide Web} - Ensuring that the validity of their 5G network access points is connected to Ref. \cite{s21} is a process that involves navigating through the evolution of technology over time.

\subsubsection{RQ2: What are approaches to improve 5G security?}
The cybersecurity of networks has changed due to the emergence of 5G. A 5G cybersecurity solution can help protect networks from unauthorized access and use. It can also be integrated with other networks to ensure maximum protection. The development of artificial intelligence and other cutting-edge technologies like the internet of things and cloud computing can also be supported by 5G security. It can be done through the use of network virtualization and deep packet inspection. DPI is a type of inspection that carefully monitors the packets sent and received over a computer network \cite{s27}. 5G is expected to have a powerful impact on cybersecurity due to its ability to help identify and prevent cyber threats. It can also help organizations improve their data analysis and communication capabilities. With the ability to integrate with smart devices, cyber audits will have a wider scope to address vulnerabilities \cite{s27}.
In RQ1, the researchers discussed the various applications that will be able to benefit from the 5G networks. These include data networks, storage, and security. They also explored how the technology can be used to enhance navigation networks. 

\textbf{IoT} - The lack of a secure and dependable communication network has hampered the advancement of the Internet of Things (IoT) due to the lack of standardization and the complexity of the network. Organizations will find it simpler to implement the essential solutions to satisfy their business objectives once 5G is available \cite{s22}. Although the various service requirements for IoT are not covered in this paper, the security and performance-related aspects are relevant nonetheless. Some of the basic capabilities that are needed for the development of IoT are summarized in Ref. \cite{s24}. Bio-inspired algorithms are also provided in Ref. \cite{s18}.

\textbf{Data storage and sharing} - The increasing number of devices and data interactions that will be enabled by 5G could threaten the consumer's trust in the organizations and devices that use it. This could also lead to the development of new privacy and security concerns \cite{s28}.

\textbf{Network security} - The goal of this project is to provide a secure 5G network in the Internet of Things environment. It will also improve the performance of wireless network communication by replacing the traditional methods of securing the data dimension with a software-defined network security structure. In terms of attack prevention, this method will use a new type of encryption known as ciphertext to replace the traditional plaintext information \cite{s18}.
The researchers developed a software-defined security structure that can detect an attack signal based on the differences and similarities of equipment manufacturing. This method can prevent unauthorized access to the sensitive data that's stored in the system. They claim that by preventing the centralized exposure of the data, the technology can enhance the security and functionality of the 5G network. The researchers observed that by defending the current network architecture from attacks, the software-defined security structure can stop illegal access to the 5G network. It is also capable of identifying assaults committed in an IoT context Ref. \cite{s25}.

\textbf{Privacy data user} - Due to the nature of the IoT environment, the data collected by various devices and services will be vulnerable to various attacks. These include distributed denial of service (DDoS) and other attacks Ref. \cite{s20}.
Big data collected by the IoT will be used for various purposes, such as analyzing and storing it. It will also be used to provide various services and improve the performance of the network. Unfortunately, the data collected by the IoT will be accessible by multiple entities and become unowned by the owners. Due to the nature of the IoT environment, the privacy of the data collected by the devices and services will be at risk. To minimize unauthorized access to the data, service providers will have to provide effective identity management solutions. IMSI capturing, a sort of deception that uses phony base stations to deceive users into submitting their IMSI, is one of the most frequent attacks that may be made against the privacy of the data \cite{s20}.

\textbf{Usability and navigation of the World Wide Internet} - In order to enhance the security and performance of the 5G network, it will be used to monitor and store the data related to the access control points Ref. \cite{s21}.

\subsubsection{RQ3: What are the methods available to improve 5G security?}
The primary studies that were conducted during the research process on 5G security were included in the RQ1 and RQ2 papers. They provide the necessary information to help the researchers develop effective strategies to enhance the system.

The intent of the studies was to provide a comprehensive overview of the various steps involved in the development of 5G networks Ref. \cite{s19}. They also analyzed the potential threats that could affect the system's operations. The distributed DoS is one of the most significant elements that could impact the system's security. The distributed DoS is one of the most significant elements that could impact the system's security. The introduction of adversarial attacks is one of the most frequent dangers that can have an impact on the system's functionality. In AI-based applications, this type of attack can be used to prevent the data from being stored in the learning phase. To prevent this, a secure system should have the necessary tools to detect and prevent these attacks in Ref \cite{s30}.
\section{Future research directions for 5G security }\label{future}
The research directions of this literature review will be used to develop a more comprehensive study of the security issues that are related to 5G security.
Researchers believe that improving the security of 5G networks can help prevent unauthorized access and exploitation \cite{a22}. Some of the areas that could be addressed in the future include developing effective and efficient malware detection techniques.
Unfortunately, most of the security solutions that are proposed for 5G networks have not been thoroughly tested and implemented in the field. One of the most common tools that researchers use to design and implement these solutions is the use of a field-programmable gate array (FPGA) \cite{a21}. In this study \cite{a23}, the authors proposed a secretory scheme that can be used to protect the data stored in a 5G network.
\begin{itemize}

 \item {\textbf{SDMN}} - One of the most critical factors that researchers need to consider when it comes to developing and implementing security solutions for 5G networks is the availability of new security schemes \cite{a31}. As the technology is based on multiple software-defined networking controllers, it is necessary to ensure that the security policies are synchronized across all of them \cite{a32}. One of the most effective ways to ensure that the security policies are enforced across the various interfaces of the software-defined networking (SDN) is through IPsec tunneling.

 \item {\textbf{Integrating artificial intelligence}} - As the technology is based on multiple software-defined networking controllers \cite{a33}, \cite{a34}, it is necessary to ensure that the security policies are synchronized across all of them \cite{a35}. One of the most effective ways to prevent unauthorized access and exploitation is by implementing security techniques that are based on artificial intelligence. Artificial intelligence could be used to improve the security of networks by allowing them to perform various tasks such as authentication and authorization. Whereas, security-based solutions are mainly used to implement techniques that are related to the detection of anomalous activities. Due to the increasing number of AI-enabled hacking attacks, more advanced security techniques are needed to combat these threats \cite{a36}.

 \item {\textbf{Network Slicing}} - One of the security concerns that emerged during the development of 5G networks is network slicing. Researchers have been studying this issue in detail \cite{a24}. A secure network slice is required to maintain the security of the network services. Several tactics have been suggested to reduce the network's vulnerability to security breaches \cite{a25}. The network slice is an independent component that carries all the necessary services and communications related to the security of the network. It can be used to implement various security-related functions \cite{a37}. With this solution, there is always a dedicated resource for security.

 \item {\textbf{Communication channels/interfaces}} - In order to implement secure communication channels in 5G, various security protocols are required to be modified. These include cryptography protocols, radio frequency fingerprinting, adaptive security schemes, and physical layer security \cite{a38}. Existing security solutions are prone to experiencing various limitations such as high overhead, lack of coordination, and resource consumption. Due to the nature of the communication channels in 5G, new security strategies are needed to address the various security issues that are expected to arise in the future \cite{a26}.

 \item {\textbf{Privacy of stored data }}- The proper management of the data that is gathered and stored in 5G networks is one of the most crucial aspects when it comes to adopting security policies \cite{a39}, \cite{a40}. This is especially important in the development of smart healthcare applications \cite{a27}. In today's privacy-centric environment, health monitoring devices are being used to monitor a patient's health conditions \cite{a41}. The data collected and stored by these devices are then sent to the cloud servers for further processing and storage. The nature of the communication environment that is used in these applications can be affected by various factors, such as the availability of resources and the potential attackers \cite{a28},\cite{a29}, \cite{a30}. This can lead to the leakage of the data at transit and the unauthorized access to the stored data. To minimize the impact of security breaches, new protocols are needed to be developed.

 \item {\textbf{Security as a Service}} - The development of 5G networks is expected to require the availability of reliable and secure security solutions to meet the needs of various industries, such as transportation, healthcare, and smart cities \cite{a42}. To meet the requirements of these organizations, various security services are proposed to be offered through a Software-as-a-service model. A number of security services, including identification, penetration testing, and monitoring, can be offered using this type of strategy \cite{a43}.
\end{itemize}
Therefore, there will be a huge field for academics to explore on the security end and how to construct better and stronger structures for its security as technology advances and in the near future switches from 5G to 6G.

\section{Conclusion}\label{conc}
The purpose of this systematic literature analysis is to locate the many study findings that have been made public about the security concerns around 5G networks. The initial keyword searches indicate that the networks are expected to provide various services such as transportation, health, and smart city. Due to the nature of the data collected and stored in these networks, the security of these networks is very important.
The review started by talking about the evolution of the mobile generation and the 5G and 4G networks. Then, it concentrated on the 5G network security challenges. In addition, the discussion covered the different meanings of cybersecurity and information security. Furthermore, the research findings by the authors of this review were presented. These included the work of Q.Tang \cite{a9}, A.Dutta \cite{a18}, and Ahmad I \cite{a15}.
\\We anticipate the 5G networks to be considered a game-changer in the telecommunication industry due to their ability to support massive bandwidth and their ability to provide reliable and low-latency communication. However, they also have a wide range of security issues that need to be addressed. This is why it is important that the security of these networks is continuously improved.
\\The review mainly focused on cyber security \cite{a52}, \cite{a53}. Due to the evolution of the 5G network's landscape, it has become more common for security threats to appear at different levels. Through a systematic literature review, the authors of this review were able to provide a comprehensive understanding of the numerous security risks with 5G networks. There are numerous security risks with 5G networks.  The review mainly focused on cyber security. Due to the evolution of the 5G network's landscape, it has become more common for security threats to appear at different levels. Through a systematic literature review, the authors of this review were able to provide a comprehensive understanding of the numerous security concerns surrounding 5G networks.
\\The authors of this review also highlighted the various opportunities that are available for the development of new research programs related to the security issues related to the 5G networks. In addition, they have presented the necessary research directions that will be needed to implement the security issues related to the 5G networks.

\bibliographystyle{IEEEtran}
\bibliography{Ref}
\end{document}